\documentclass[aps,prl,reprint,graphicx,superscriptaddress]{revtex4-1}

\usepackage{color}
\usepackage{amsmath}
\usepackage{graphicx}
\usepackage{multirow}
\usepackage{amsfonts}
\usepackage{amssymb}
\usepackage{amscd}
\usepackage{multirow}
\usepackage{dcolumn}% Align table columns on decimal point
\usepackage{bm}% bold math
\usepackage{float}
\usepackage{framed}

\newcommand{\bra}[1]{\mbox{$\left\langle #1 \right|$}}
\newcommand{\ket}[1]{\mbox{$\left| #1 \right\rangle$}}

\begin{document}

\title{Quantum Coherence Witness with Untrusted Measurement Devices}

\author{You-Qi Nie}
\affiliation{Hefei National Laboratory for Physical Sciences at Microscale and Department
of Modern Physics, University of Science and Technology of China, Hefei, Anhui 230026, China}
\affiliation{CAS Center for Excellence in Quantum Information
and Quantum Physics, University of Science and Technology of China, Hefei, Anhui 230026, China}

\author{Hongyi Zhou}
\affiliation{Center for Quantum Information, Institute for Interdisciplinary Information Sciences,
Tsinghua University, Beijing 100084, China}

\author{Jian-Yu Guan}
\affiliation{Hefei National Laboratory for Physical Sciences at Microscale and Department
of Modern Physics, University of Science and Technology of China, Hefei, Anhui 230026, China}
\affiliation{CAS Center for Excellence in Quantum Information
and Quantum Physics, University of Science and Technology of China, Hefei, Anhui 230026, China}

\author{Qiang Zhang}
\affiliation{Hefei National Laboratory for Physical Sciences at Microscale and Department
of Modern Physics, University of Science and Technology of China, Hefei, Anhui 230026, China}
\affiliation{CAS Center for Excellence in Quantum Information
and Quantum Physics, University of Science and Technology of China, Hefei, Anhui 230026, China}

\author{Xiongfeng Ma}
\email{xma@tsinghua.edu.cn}
\affiliation{Center for Quantum Information, Institute for Interdisciplinary Information Sciences,
Tsinghua University, Beijing 100084, China}

\author{Jun Zhang}
\email{zhangjun@ustc.edu.cn}
\affiliation{Hefei National Laboratory for Physical Sciences at Microscale and Department
of Modern Physics, University of Science and Technology of China, Hefei, Anhui 230026, China}
\affiliation{CAS Center for Excellence in Quantum Information
and Quantum Physics, University of Science and Technology of China, Hefei, Anhui 230026, China}

\author{Jian-Wei Pan}
\affiliation{Hefei National Laboratory for Physical Sciences at Microscale and Department
of Modern Physics, University of Science and Technology of China, Hefei, Anhui 230026, China}
\affiliation{CAS Center for Excellence in Quantum Information
and Quantum Physics, University of Science and Technology of China, Hefei, Anhui 230026, China}

\date{\today}

\begin{abstract}
Coherence is a fundamental resource in quantum information processing, which can be certified by a coherence witness. Due to the imperfection of measurement devices, a conventional coherence witness may lead to fallacious results. We show that the conventional witness could mistake an incoherent state as a state with coherence due to the inaccurate settings of measurement bases. In order to make the witness result reliable, we propose a measurement-device-independent coherence witness scheme without any assumptions on the measurement settings. We introduce the decoy-state method to significantly increase the capability of recognizing states with coherence. Furthermore, we experimentally demonstrate the scheme in a time-bin encoding optical system.
\end{abstract}

\maketitle %\maketitle must follow title, authors, abstract and \pacs

%\section{introduction}
%\emph{Introduction.---}
Superposition explains many striking phenomena of quantum mechanics, such as the interference in the double-slit experiment of electrons and Schr\"{o}dinger's cat gedanken experiment. According to Born's rule, measuring a superposed system would lead to a random projection, whose outcome cannot be predicted in principle. This feature can be employed in quantum information processing for designing quantum random number generators (QRNGs) \cite{MaQRNG,RevModPhys.89.015004}. Recently, the strength of superposition is quantified under the framework of quantum coherence \cite{baumgratz2014quantifying,RevModPhys.89.041003}, which is a rapidly developing field in quantum foundation. Quantum coherence has close connections with entanglement and other quantum correlations in many-body systems, and interestingly these measures can be transformed into each other \cite{PhysRevLett.115.020403,PhysRevLett.116.160407,yuan2017unified,PhysRevA.99.022326}. Also, various concepts can be mapped from quantum entanglement to quantum coherence, such as coherence of assistance \cite{chitambar2016assisted}, coherence distillation and cost \cite{yuan2015intrinsic,winter2016operational,Zhao2018OneShot,PhysRevLett.121.010401,zhao2019one}, and coherence evolutions \cite{PhysRevA.89.024101}. It turns out that coherence, as an essential resource, plays an important role in various tasks including quantum algorithms \cite{hillery2016coherence}, quantum biology \cite{o2014non}, and quantum thermodynamics \cite{goold2016role}.

In reality, it is crucial to judge whether a quantum source is capable for certain quantum information processing tasks. Coherence witness has been introduced to detect the existence of coherence for an unknown state \cite{napoli2016robustness}. A valid coherence witness $W$ is a Hermitian operator which is positive semidefinite after dephasing on the coherence computational basis $\Delta(W)\geq 0$. This condition is equivalent to that of $\mathrm{tr}(\rho W)\geq 0$ for all incoherent states. Then, $\mathrm{tr}(\rho W)<0$ shows coherence in $\rho$.
Coherence witness has a close connection with a coherence measure called robustness of coherence $C_{\mathcal{R}}(\rho)$ \cite{napoli2016robustness}. If we optimize the observable $W$ to maximize $-\mathrm{tr}(\rho W)$, the maximum value is the robustness of coherence of $\rho$. In other words, the witness can be used to lower bound the coherence of an unknown system \cite{PhysRevX.8.041007}; i.e., the relation $C_{\mathcal{R}}(\rho)\geq -\mathrm{tr}(\rho W)$ always holds for a valid witness $W$ \cite{napoli2016robustness}. This property can also be applied to construct a source-independent QRNG \cite{ma2017source}. Several experiments relevant to coherence witness have been reported recently \cite{PhysRevLett.118.020403,PhysRevLett.120.230504,PhysRevX.8.041007}.

The key problem is that the correctness of coherence witness highly relies on the implementations of $W$, whose results may be unreliable due to measurement device imperfections or malfunction. As an example, we propose a simple basis-rotating attack (as a way to mimic device malfunction) on the measurement devices. As a result, an incoherent state is mistaken for a state with nonzero coherence. Considering the $Z$-basis coherence witness $W^0=1/2+\sigma_x/2+\sigma_z/2$, we can easily check $\mathrm{tr}(\rho W^0)>0$ for all incoherent states $\rho=p\ket{0}\bra{0}+(1-p)\ket{1}\bra{1}$ in the $Z$ basis. However, if the adversary rotates the measurement setting of $\sigma_x$ to $\sigma_z$, the actual witness becomes $W^1=1/2+\sigma_z$, which leads to an incorrect witness when $p<1/4$ (see Section I in Supplemental Material~\cite{sm} for detailed discussions, which includes Refs.~\cite{kurotani2007upper,zorzi2014minimum,coles16,ma2005practical,Ma13,NIST}).

This would lead to serious consequences in practice. In the case of QRNG implementation, where the source entropy is characterized by coherence witness, the unreliable results can bring security loopholes for its cryptographic applications. Similarly, a wrong estimation of coherence can also result in poor success probabilities \cite{anand2016coherence} or precisions of quantum algorithms \cite{matera2016coherent}.

In this Letter, we propose a measurement-device-independent coherence witness (MDICW) that is robust against any bias on measurement devices, inspired by the measurement-device-independent entanglement witness (MDIEW) scheme addressing the detection imperfection in entanglement witness \cite{branciard2013measurement,xu2014implementation}. The main differences between the two schemes are compared in Table~\ref{tab1}. Compared with the conventional coherence witness that requires complete characterization and manipulation of the measurement devices, our MDICW method can remove the requirements on the measurement device and need only one measurement setting, which provides a stronger tool to detect and lower-bound coherence in an unknown system.

\begin{table}[htbp]
\centering
\caption{Comparison between entanglement witness (EW) and coherence witness (CW). $\omega_t$ and $\tau_s$ are test quantum states, $a$ and $b$ are classical outputs, and $\beta_{a,b}^{s,t}$ is a real coefficient in MDIEW. QKD: quantum key distribution.}
\begin{tabular}{ccc}
    \hline
    Task & EW &CW \\
    \hline
    Common criteria & $\mathrm{tr}(\rho W_E)$ & $\mathrm{tr}(\rho W)$ \\
    MDI criteria &   $\underset{a,b,s,t}\sum\beta_{a,b}^{s,t}p(a,b|\omega_t, \tau_s)$  & Eq.~\eqref{eq:targetfunc} \\
    Inspiration & MDI-QKD \cite{PhysRevLett.108.130503} & MDI-QRNG \cite{cao2015loss} \\
    \hline
\end{tabular}
\label{tab1}
\end{table}

Following the idea of MDI-QRNG \cite{cao2015loss,Nie16}, we perform tomography of the untrusted measurement where the test states are chosen to be eigenstates of Pauli matrices. The coherence of an unknown state can be lower bounded by the tomography results. In practice, since weak coherent states are used as approximations of ideal qubit states, there are inevitable deviations in the tomography results and the coherence lower bound can be quite loose, which makes it difficult to identify states with coherence. In other words, the coherence in most states cannot be detected.
%Specific to QRNG that extracts coherence as randomness, it results in that the guaranteed lower bound of randomness is far lower than the actual value, and thus considerably limits the generation rate of QRNG.
To deal with this issue, the decoy state method from quantum key distribution \cite{Hwang2003Decoy,lo2005decoy,Wangxb2005decoy} is introduced to tighten the lower bound of coherence. To show the improvement, we make a comparison between the cases with and without a decoy state method.
Besides the main scheme of MDICW, we also design a control experiment where we mix two coherent states and observe the vanish of coherence, showing the convexity of coherence.

\begin{figure}[tbp]
\begin{minipage}[b]{1\linewidth}
\centering
\includegraphics[width=8.5 cm]{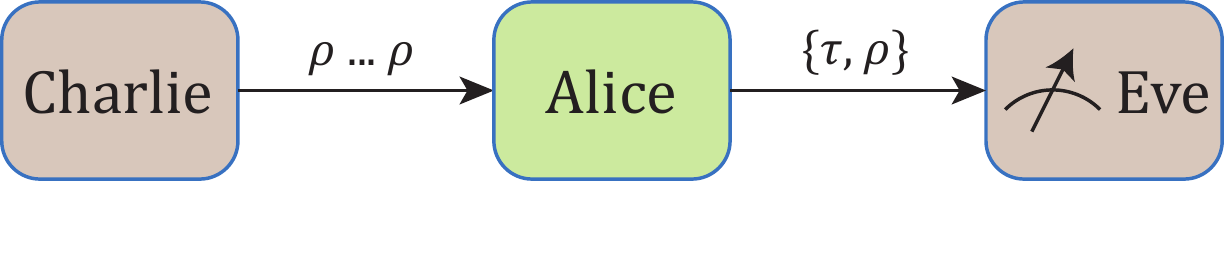}
\end{minipage}
\begin{minipage}[b]{1\linewidth}
\centering
\begin{framed}
\centering
\begin{enumerate}
\item
Charlie prepares qubit state $\rho$ unknown to Alice.
\item
Alice prepares her test states from a set $\{\tau\}$, so she constitutes an expanded states set $\{\tau,\rho\}$.
\item
Alice randomly sends the states from the set $\{\tau,\rho\}$ to an untrusted measurement device.
%\item Alice prepares her test states from a set $\{\tau\}$, which are mixed with $\rho$ and sent to Eve (an untrusted measurement device), Eve randomly receives test state from the set of $\{\tau\}$ and unknown who announces her binary measurement results.
\item
Alice records the loss events and double click events to be $0$ and calculates the conditional probabilities $p(1|j)$ ($j\in \{\tau,\rho\}$).
\item
Alice calculates a lower bound of coherence of $\rho$ on a certain basis with Eq.~\eqref{eq:targetfunc}. If the lower bound is nonpositive, no coherence is witnessed.
\end{enumerate}
\end{framed}
%\captionof{table}{Table}
\caption{MDICW scheme.}\label{fig1}
\end{minipage}
\end{figure}

The MDICW scheme works as follows. An untrusted party, Charlie, prepares independent and identically distributed unknown state $\rho$. These states are sent to Alice, who wants to detect coherence in $\rho$ in a given computational basis and certify the lower bound of coherence. Alice prepares some test states from a set $\{\tau\}$ to make a tomography of the untrusted measurement designed by Eve. Here, we assume that $\{\tau\}$ and $\rho$ are in the same support. The measurement site would randomly receive a test state from $\{\tau\}$ or the unknown state $\rho$. In our implementation, the set of test states $\{\tau\}$ are chosen to be eigenstates of Pauli matrices $\{\ket{0},\ket{1},\ket{+},\ket{+i}\}$ for simplicity. After receiving the states, Alice could obtain measurement results, $0$, $1$, loss, and double click. Alice records the loss and double click events to be $0$, which makes the scheme loss tolerant \cite{cao2015loss}. Then Alice calculates the probabilities of output $1$ conditioned on different input states $p(1|j)$ ($j\in \{\tau,\rho\}$) to get the tomography result of a qubit POVM $M_0$ and $M_1$. Eventually, Alice can evaluate the coherence lower bound. The protocol is summarized in Fig.~\ref{fig1}.

First, we consider an ideal case where the test states $\{\ket{0},\ket{1},\ket{+},\ket{+i}\}$ are perfect qubits. Then, the tomography result is a qubit POVM uniquely determined by a set of parameters $\{a_1,n_x,n_y,n_z\}$ \cite{kurotani2007upper},
\begin{equation}
\begin{aligned}
M_0 & = I- M_1 \\
M_1 & = a_1(I+n_x \sigma_x+n_y \sigma_y + n_z \sigma_z),
\end{aligned}
\end{equation}
where $I$ is the two-dimensional identity matrix. The conditional probabilities are given by
\begin{equation}
\begin{aligned}
p(1\big|\ket{0}\bra{0})&=a_1+a_1n_z, \\
p(1\big|\ket{1}\bra{1})&=a_1-a_1n_z, \\
p(1\big|\ket{+}\bra{+})&=a_1+a_1n_x, \\
p(1\big|\ket{+i}\bra{+i})&=a_1+a_1n_y, \\
\end{aligned}
\end{equation}
where $n_x^2 +n_y^2 +n_z^2  \leq 1$ and $0 \leq a_1 \leq 1$. With the measurement conditional probabilities $p(1|j)$ ($j\in \{\ket{0},\ket{1},\ket{+},\ket{+i}\}$), Alice can make a full tomography of the qubit POVM $\{M_0,M_1\}$. One can refer to Section II in Supplemental Material for details~\cite{sm}.

Further, we try to find the coherence lower bound of the unknown state $\rho$ given the tomography result, which is a convex optimization problem by minimizing the relative entropy measure of coherence \cite{baumgratz2014quantifying},

%\section{Coherence witness}
%\emph{Coherence witness as a convex optimization problem}

\begin{equation}
\min_{\rho} C_{rel.}(\rho)=\min_{\rho} \min_{\sigma \in \mathcal{I}}S(\rho||\sigma)
\end{equation}
with the constraint of
\begin{equation}
P(1|\rho) = \mathrm{tr}(\rho M_1),
\end{equation}
where $\mathcal{I}$ is the set of incoherent states $\sigma=\sum_i p_i\ket{i}\bra{i}$ on the computational basis $\{\ket{i}\}$.
The primal problem can be transformed into a dual problem \cite{zorzi2014minimum,coles16}
\begin{equation}\label{eq:targetfunc}
\max_\lambda[-||\sum_i \Pi_i \exp(-\mathbb{I}-\lambda M_1) \Pi_i ||-\lambda \mathrm{tr}(\rho M_1)],
\end{equation}
where the infinity norm is to find the maximum eigenvalue of the matrix, $\Pi_i$ is the projective measurement corresponding to the computational basis $\{\ket{i}\}$, and $\mathbb{I}$ is the identity matrix (see Section \uppercase\expandafter{\romannumeral 3} in the Supplemental Material~\cite{sm} for the details).

\begin{figure*}[tbp]
\centering
\includegraphics[width=16 cm]{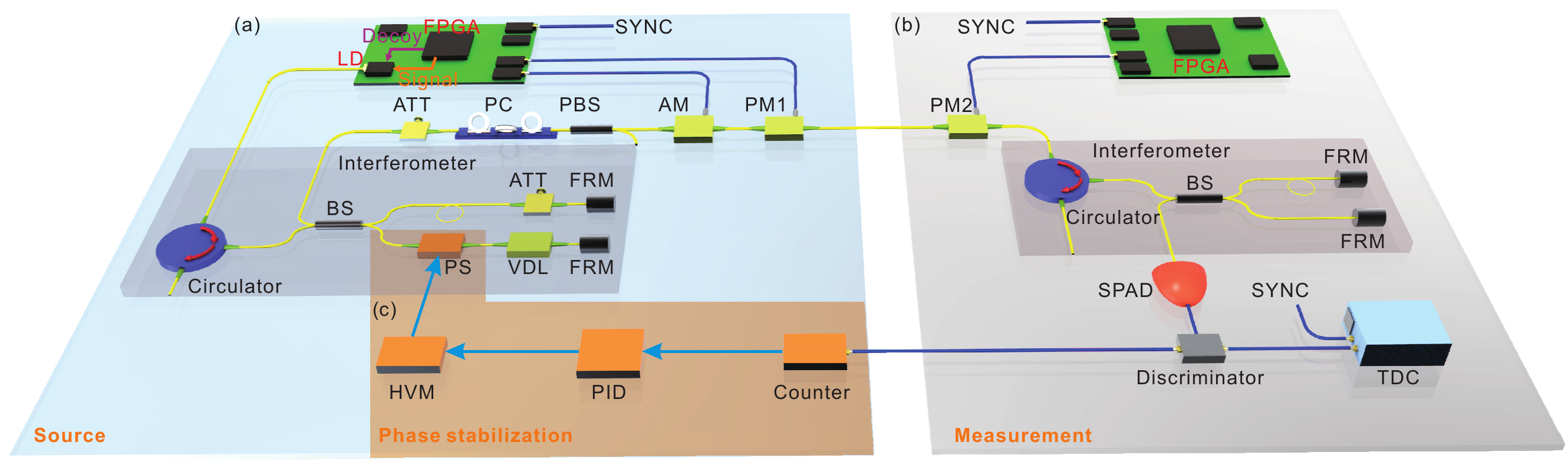}
\caption{Experimental setup for the MDICW scheme including the source part (a), the measurement part (b), and the phase stabilization part (c).
LD: laser diode, FPGA: field-programmable gate array, SYNC: synchronized signal, BS: beam splitter, ATT: attenuator, FRM: Faraday rotator mirror, PS: phase shifter, VDL: variable delay line, HVM: high-voltage module, PID: proportional-integral-derivative algorithm, PC: polarization controller, PBS: polarizing beam splitter, AM: amplitude modulator, PM: phase modulator, SPAD: single-photon avalanche diode, TDC: time-to-digital converter.}
\label{fig2}%
\end{figure*}

%\emph{Applying decoy state method}
In practice, phase randomized weak coherent states are widely used as approximations of single-photon sources, which leads to biases in the tomography result; i.e., we can only get some bounds on the set of parameters $\{a_1,n_x,n_y,n_z\}$ rather than their accurate values. For each value of the conditional probability recorded by Alice, it may come from different photon number components
\begin{equation}
p_{\mu}(1|j)=e^{-\mu}\sum_n \frac{\mu^n}{n!}p_n(1|j),
\end{equation}
where $\mu$ is the mean photon number of the signal state.
What we care about is the single photon component contribution $p_1(1|j)$ ($j\in \{\ket{0},\ket{1},\ket{+},\ket{+i},\rho\}$) in our tomography. To estimate the value of $p_1(1|j)$ more accurately, we apply the decoy state method, i.e., by adjusting the intensities of input states.
It has been proven that vacuum and weak decoy states are enough to estimate $p_1(1|j)$ \cite{ma2005practical},
\begin{equation}\label{eq:constraint}
\begin{aligned}
\frac{\mu}{\mu \nu -\nu^2}\left(p_{\nu}(1|j)e^\nu-p_{\mu}(1|j)\frac{\nu^2}{\mu^2}-\frac{\mu^2-\nu^2}{\mu^2}p_d\right) \\
\leq p_1(1|j) \leq \frac{p_\nu(1|j)}{\nu e^{-\nu}},
\end{aligned},
\end{equation}
where $p_d$ is the dark count rate of detector estimated by the vacuum state, and $\nu$ is the mean photon number of the weak decoy state.
Then, the lower bound of the relative entropy measure of coherence can be obtained by optimizing Eq.~\eqref{eq:targetfunc} with constraints of Eq.~\eqref{eq:constraint}. We compare the performances of MDICW with and without the decoy state method at the end of this Letter.

Here are some remarks about the protocol. First, we assume that the unknown state is on the same support of the test states. This assumption comes from the squashing model in the security analysis of quantum communication \cite{PhysRevLett.101.093601}, where the tomography result, the two-output POVM, is just an effective POVM in the subspace of the test states $\{\tau\}$. We can always squash the unknown state into the subspace of $\{\tau\}$ and calculate the coherence lower bound of the squashed input state with our MDICW method. Since the squasher can be incoherent operations in the computational basis, the lower bound also holds for the original input state $\rho$. Second, in conventional coherence witness reported in literature \cite{PhysRevLett.118.020403,PhysRevLett.120.230504,PhysRevX.8.041007}, usually multiple measurement settings are required, e.g., $W=a\sigma_x+b\sigma_z$ ($a$ and $b$ are real coefficients), and the coherence lower bound is given by $-\mathrm{tr}(\rho W)$. While in our protocol, we only use one measurement setting with multiple state preparations. In fact, there is only a $Y$ basis measurement in our experiment. The lower bound is based on the uncertainty relation of conjugate measurement basis intuitively. Third, we also apply the decoy state method to the unknown state $\rho$ to get the constraints in Eq.~\eqref{eq:constraint}. This is because the quantum states are characterized in the degree of freedom of polarization or phase, rather than intensity. Alice can control the intensity and insert an attenuation before it is detected, which can effectively be regarded as the decoy state method. Of course, one can also get a looser lower bound without the decoy state method.
%\section{Experiment}
%\emph{Experiment}

Furthermore, we experimentally demonstrate the MDICW scheme with decoy state method using a time-bin encoding system, and Fig.~\ref{fig2} illustrates the experimental setup.
The required quantum states in $X$, $Y$, and $Z$ bases are randomly prepared with different intensities in the source part, and real-time active basis switch is performed in the measurement part.

In the source part, as shown in Fig.~\ref{fig2}(a), a 1550 nm laser diode (LD) is driven by narrow pulses with different amplitudes to create phase-randomized laser pulses with different intensities, corresponding to signal states and decoy states, respectively. The laser pulses enter an unbalanced interferometer with a time delay of $\sim$ 4.8 ns to form two time-bin pulses.
The output pulses from the interferometer pass through in sequence a tunable (ATT), a polarization controller (PC), and a polarizing beam splitter (PBS).
The output of PBS is further modulated by two polarization-maintaining components, i.e., an amplitude modulator (AM) and a phase modulator (PM1), which are controlled by a field-programmable gate array (FPGA). With such configuration, all required time-bin quantum states can be prepared in real-time.

In the measurement part, as shown in Fig.~\ref{fig2}(b), the incident photons are further modulated by PM2 controlled by another FPGA and then enter into another interferometer
that has the same time delay as that in the source part. The output photons from the interferometer are detected by an InGaAs/InP single-photon avalanche diode (SPAD) with 1.25 GHz sine wave gating~\cite{Liang12}. Different pulse amplitudes for the modulation of PM2 are used to perform $X$ or $Y$ basis measurements.

In the experiment, in order to implement the phase stabilization between two interferometers and the channel transmission loss as low as possible, a variable delay line (VDL) and a phase shifter (PS) are inserted into one arm of the interferometer in the source part,
and active feedback technology is applied by precisely tuning the PS in real-time for phase stabilization (see Section \uppercase\expandafter{\romannumeral 5} in Supplemental Material~\cite{sm} for the details of phase stabilization).
Considering $25\%$ detection efficiency of the SPAD, the insertion losses of PM2 and interferometer, the total transmission efficiency $\eta$ of the system is $\sim 4.86\%$, corresponding to a loss of about 13.13 dB.

The quantum states of $\ket{0}$, $\ket{1}$, $\ket{+}$, $\ket{+i}$, $\ket{-}$, and $\ket{-i}$ are prepared and verified carefully. Typical count rate distributions of the six time-bin states are measured in $X$, $Y$, and $Z$ bases using SPAD and TDC. To implement the  $Z$ basis measurement, PM2 and the interferometer in the measurement part are not used. For $X$ ($Y$) basis measurement, the relative phase between two pulses is set as 0 ($\frac{\pi}{2}$) by PM2.
Further, we measure the error rates of the prepared states after the projection in $X$, $Y$, and $Z$ bases, respectively. The average values of error rates are pretty low with slight fluctuations, which indicates the accuracy and stability of the quantum state preparation. The error rates are mainly attributed to the optical misalignment, the dark counts and afterpulses~\cite{ZIZ15} of the InGaAs/InP SPAD (see Section \uppercase\expandafter{\romannumeral 5} in the Supplemental Material~\cite{sm} for the details).

\begin{table}[tbp]
\centering
\tabcolsep0.05in
\caption{Results of measurement tomography.
\label{table2}}
\begin{tabular}{lc|ccc}
  \hline
  \multicolumn{2}{c}{Test state}             & Amount   & Counts of ``1'' & Probability           \\
  \hline
  \multirow{4}{*}{Signal state}& $\ket{0}$     & 2049836  & 21671         & $1.06\times10^{-2}$   \\
                               & $\ket{1}$     & 2049204  & 24354         & $1.19\times10^{-2}$   \\
                               & $\ket{+}$     & 2047279  & 22753         & $1.11\times10^{-2}$   \\
                               & $\ket{+i}$    & 2048073  & 45306         & $2.21\times10^{-2}$   \\
                               & $\rho$        & 8188952  & 182115        & $2.22\times10^{-2}$   \\
%  \cline{2-5}
  \hline
  \multirow{4}{*}{Decoy state} & $\ket{0}$     & 2046756  & 2303          & $1.13\times10^{-3}$   \\
                               & $\ket{1}$     & 2047612  & 2467          & $1.20\times10^{-3}$   \\
                               & $\ket{+}$     & 2049153  & 2464          & $1.20\times10^{-3}$   \\
                               & $\ket{+i}$    & 2048549  & 4517          & $2.20\times10^{-3}$   \\
                               & $\rho$        & 8192586  & 18497         & $2.26\times10^{-3}$   \\
  \hline
\end{tabular}
\end{table}

During the experiment, the four time-bin quantum states of $\ket{0}$, $\ket{1}$, $\ket{+}$, $\ket{+i}$, and an unknown state $\rho$ with intensities of $\mu$ or $\nu$ are randomly sent, while the measurement part is randomly chosen between $X$ and $Y$ bases. Without loss of generality, the unknown quantum state is set as $\ket{+i}$ and the unknown measurement for MDICW process is set as $Y$ basis measurement.

The number of prepared states to perform coherence witness is $3.3 \times 10^{7}$. The measurement tomography results are listed in Table~\ref{table2}. By applying the evaluation method of coherence witness, the coherence of the unknown state $\rho$ is lower bounded by $0.25$ per detected signal state.

\emph{Control experiment.---} In order to verify the effectiveness of the MDICW scheme, a control experiment is designed and performed using the same experimental setup. The four time-bin test states of $\ket{0}$, $\ket{1}$, $\ket{+}$, $\ket{+i}$, and a mixed state $\rho^\prime$ as an ensemble of $\ket{+i}$ and $\ket{-i}$ with intensities of $\mu$ or $\nu$ are randomly sent to untrusted measurement device. As a result, no coherence is witnessed for the mixed state $\rho^\prime$. However, if we can distinguish the components of $\rho^\prime$ and divide it into two parts, $\ket{+i}$ and $\ket{-i}$, the coherence of each part is lower-bounded by $0.0285$ and $0.1279$ per detected signal state, respectively. See Section \uppercase\expandafter{\romannumeral 5} in the Supplemental Material~\cite{sm} for the details of the experiment. The results show that states with little coherence or incoherent states cannot be witnessed in our scheme, and also imply the convexity of coherence since the lower bound decreases by mixing.

\begin{figure}[tbp]
\centering
\includegraphics[width=7.5 cm]{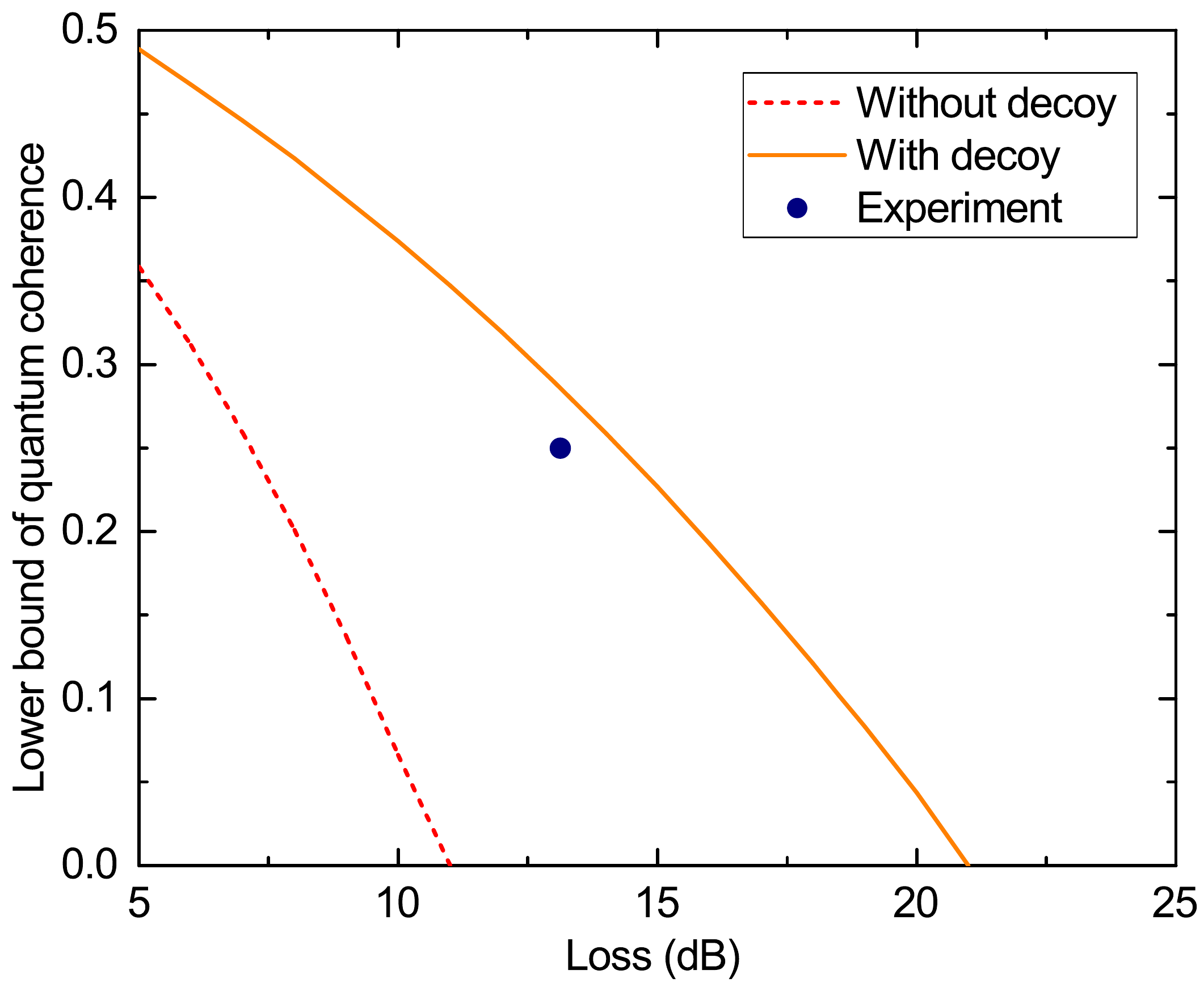}
\caption{Simulation comparison of coherence witness with (solid line) and without (dashed line) decoy state method as a function of channel loss. The simulation parameters include $p_d=10^{-6}$, $N=3.2\times10^6$, $\eta_1=\eta_2=\eta_3=\eta_4=1/8$, $\eta_5=1/2$, $p_s=0.5$, $n_\sigma$=3.89, and zero error rate. The circle point represents the experimental result with decoy state method under the condition of high channel loss of 13.13 dB and nonzero error rate.}
\label{fig:comp}%
\end{figure}

In order to show the advantage in calculating the coherence lower bound using decoy state method, we perform a simulation comparison between the two cases with and without decoy state method (see Section \uppercase\expandafter{\romannumeral 4} in the Supplemental Material~\cite{sm} for the details), as shown in Fig.~\ref{fig:comp}.
The simulation results clearly show that using decoy state method can significantly improve the quantification reliability of coherence witness and tolerate considerably high channel loss.
In the experiment, the channel loss is 13.13 dB. The conventional method without using a decoy state method even cannot quantify the coherence in such case.
In order to effectively compare with experimental results, the simulation parameters are consistent with the experimental setup except for error rate, which is hard to be precisely determined in the experiment and zero is chosen.
The experimental lower bound is a little smaller than the simulation result due to the nonzero error rate in the experiment.

In summary, we propose an MDICW scheme with the decoy state method for reliable certification of quantum coherence, and experimentally demonstrate the scheme with a time-bin encoding system. In the experiment, we obtain a lower bound of $0.25$ per detected signal state even with untrusted measurement devices. Though our protocol is inspired by the MDIEW protocol, there is a crucial difference that in MDIEW there is no dimension assumption on the unknown state. It is an interesting future direction for developing a new MDICW scheme without the dimension assumption. One possible approach is to send the unknown state together with ancillary test states to Eve, who performs an untrusted Bell state measurement to tell the fidelities between them. A similar work \cite{ma2017source} has been presented recently, where a source-independent QRNG is proposed based on the coherence witness of an unknown state. In that work, the randomness is certified by coherence witness with trusted measurement devices while in our work the measurement device is untrusted. Another difference is that in Ref.~\cite{ma2017source} results from different measurement settings ($X$, $Y$, and $Z$ basis measurement) are used to bound the coherence, whereas in our work we can only use measurement results from a single effective measurement setting (the tomography result). Also, there is a recent work on witnessing the multilevel coherence \cite{PhysRevX.8.041007} based on different assumptions. It considers the measure of robustness of coherence. While our method can deal with general coherence measures as long as they are convex.

\begin{acknowledgments}
The authors acknowledge X.~Yuan, X.~Zhang, and Q.~Zhao for helpful discussions and the technical support from the staff of QuantumCTek Co., Ltd. This work has been supported by the National Key R\&D Program of China under Grant No.~2017YFA0304004, the National Natural Science Foundation of China under Grants No.~11674307, No.~11674193, and No.~11875173, the China Postdoctoral Science Foundation under Grant No.~2018M632531, the Anhui Provincial Natural Science Foundation under Grant No.~1908085QA38, the Chinese Academy of Sciences, and the Anhui Initiative in Quantum Information Technologies.

Y.-Q.~N.~and H.~Z.~contributed equally to this work.

\end{acknowledgments}

\bibliography{bibMDICW}

\clearpage

\section{Supplemental Material: Quantum Coherence Witness with Untrusted Measurement Devices}

\section{Basis-rotating attack on conventional coherence witness}
\begin{figure}[tbph!]
\centering
\includegraphics[width=8 cm]{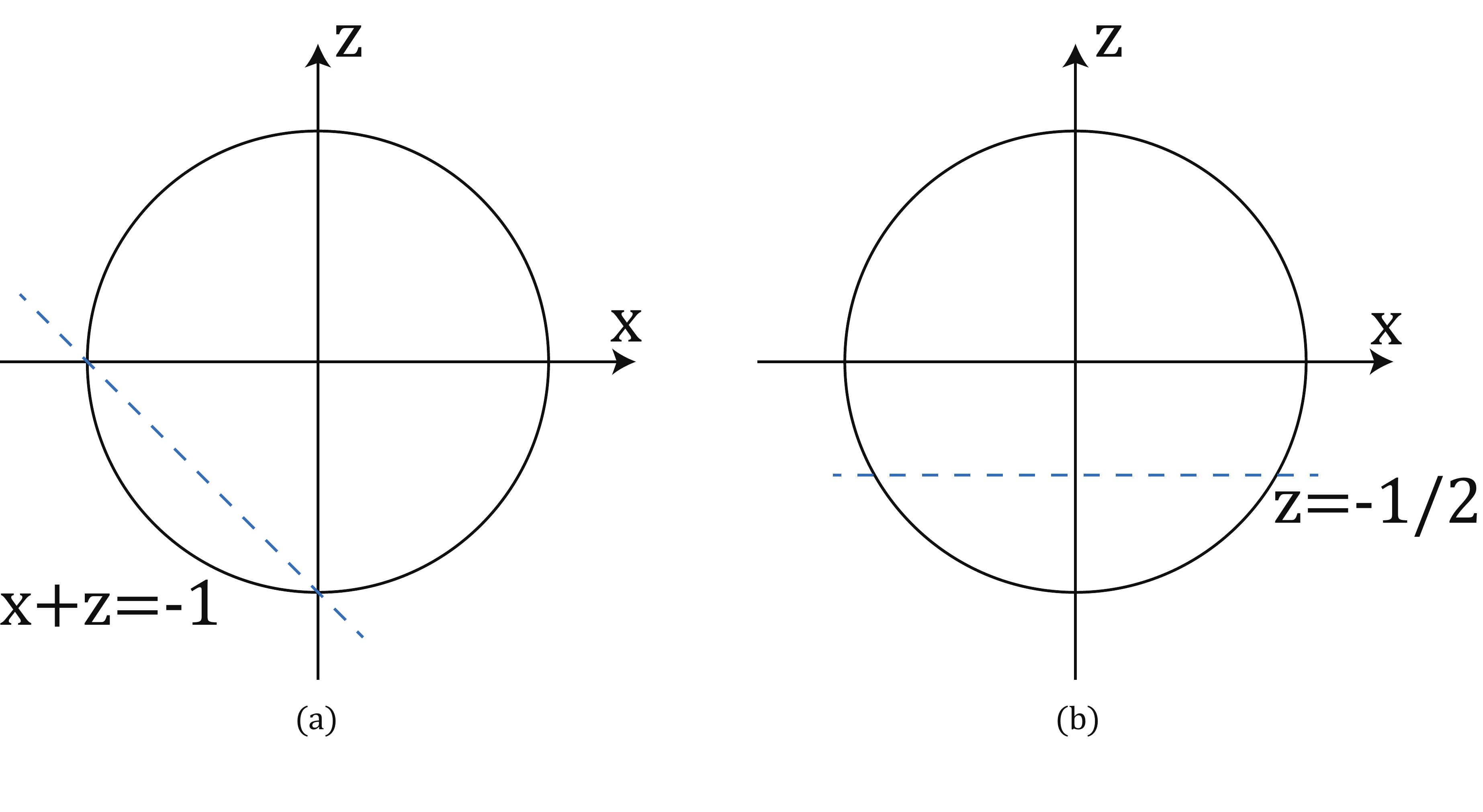}
\caption{(a) The witness. (b) The actual witness under basis rotating attack. The plane has an intersection with $Z$-axis within the Bloch sphere.}
\label{fig:attack}%
\end{figure}

The conventional coherence witness is a certain measurement $W$ that satisfies $\mathrm{tr}(\rho W)<0$ for states with non-zero coherence. Here we propose an attack, as illustrated in Fig.~\ref{fig:attack}, under which an incoherence state will be mistaken for a state with non-zero coherence when the measurement device is manipulated by an adversary, Eve. Similar to the entanglement witness, the coherence witness $W$ is a Hermitian operator. For simplicity, we consider the two dimensional $Z$ basis coherence witness. The incoherent states correspond to $Z$-axis in the Bloch sphere. The witness, $W$, has a Pauli-matrix presentation of $W=w_0I+w_1 \sigma_x + w_2 \sigma_y+w_3\sigma_z$, corresponding to a plane $w_1x+w_2y+w_3z+w_0=0$.

For a valid witness, the $Z$-axis in the Bloch sphere, i.e., $x=y=0, -1<z<1$, should be in one side of the plane. The result of the coherence witness is given by
\begin{equation}\label{eq:cw}
\mathrm{tr}(\rho W)=w_0+ \langle w_1 \sigma_x \rangle+ \langle w_2 \sigma_y \rangle+\langle w_3\sigma_z \rangle.
\end{equation}
For an arbitrary two-dimensional incoherent state $\delta=p\ket{0}\bra{0}+(1-p)\ket{1}\bra{1}$, Eq.~\eqref{eq:cw} should satisfy $\mathrm{tr}(\delta W)=w_0+w_3(2p-1)>0$. Then we can simply let $w_0=w_3$. For example, we can take a specific witness $W=1/2+w_1\sigma_x/2+w_3\sigma_z/2$. Now we consider the basis-rotating attack where Eve rotates $\sigma_x$ measurement to $\sigma_z$, then
\begin{equation}
\mathrm{tr}(\delta W)=2p-\frac{1}{2}.
\end{equation}
When $p<1/4$ the incoherent state is witnessed to be a state with non-zero coherence.

\section{Full tomography of a two-dimensional POVM}
An arbitrary qubit positive-operator valued measure (POVM) can be expressed as \cite{kurotani2007upper}
\begin{equation}
\begin{aligned}
M_0&=a_0(I+\vec{n}_0 \cdot \vec{\sigma})\\
M_1&=a_1(I+\vec{n}_1 \cdot \vec{\sigma}),
\end{aligned}
\end{equation}
where the coefficients $a_0$ and $a_1$ are real numbers, and $\vec{n_0}$ and $\vec{n_1}=(n_x, n_y, n_z)$ are real vectors satisfying
\begin{equation}
\begin{aligned}
a_0,a_1&\geq0 \\
a_0+a_1&=1 \\
|n_0|,|n_1|&\leq 1 \\
a_0\vec{n}_0+a_1\vec{n}_1&=0.
\end{aligned}
\end{equation}
The probabilities of outcome bits `0' and `1' given an input state $\rho$ are
\begin{equation}\begin{aligned}
p(0|\rho)&=tr(M_0\rho)\\
p(1|\rho)&=tr(M_1\rho).
\end{aligned}
\end{equation}
When the input states are $\ket{0}$, $\ket{1}$, $\ket{+}$ and $\ket{+i}$, the corresponding probabilities of outcome bit `1' are
\begin{equation}
\begin{aligned}\label{eq:tomography}
  p(1\big|\ket{0}\bra{0})&=a_1+a_1n_z \\
  p(1\big|\ket{1}\bra{1})&=a_1-a_1n_z \\
  p(1\big|\ket{+}\bra{+})&=a_1+a_1n_x \\
  p(1\big|\ket{+i}\bra{+i})&=a_1+a_1n_y,
\end{aligned}
\end{equation}
where $p_i$ $(i=1,2,3,4)$ can be measured in the experiment.
From Eq.~\eqref{eq:tomography}, one can find that the number of unknown parameters is equal to that of equations.
Therefore, the POVM parameters $a_1$, $n_x$, $n_y$, and $n_z$ can be calculated and the measurement tomography is accomplished.

\section{Coherence witness as a convex optimization problem}
Here we consider a problem of finding coherence lower bound of an unknown state, given the probabilities of outcomes of a certain measurement. We use the method in Ref. \cite{zorzi2014minimum,coles16} to transform such a problem into a convex optimization problem. Given an unknown qubit state $\rho$, a two-dimensional POVM $M=\{M_0, M_1\}$, and the probabilities $\mathrm{tr}(M_a \rho) =  m_a (a=0,1)$, multiple individual measurements on the qubit are performed to quantify the qubit's lower bound of coherence by the relative entropy.

The problem can be expressed as
\begin{equation}\label{eq:primaloptimization}
\begin{aligned}
    \alpha := \underset{\rho}\min\, {C_{\mathrm{rel.}}}(\rho) &=\underset{\rho}\min\,\underset{\sigma\in\mathcal{I}}\min\,S(\rho||\sigma) \\
    \mathrm{Constraints:}\mathrm{tr}(M_1 \rho) &= P(1|\rho),
\end{aligned}
\end{equation}
where $S(\rho||\sigma)$ is defined as $S(\rho||\sigma)=\mathrm{tr}[\rho log(\rho)]-\mathrm{tr}[\rho log(\sigma)]$, and the probability of result $\mathrm{tr}(M_1 \rho) = 1- \mathrm{tr}(M_0 \rho)$ is derived from the completeness condition of a POVM. For a certain computational basis $\{i\}$ of a projector $\Pi$, the optimization problem of Eq.~\eqref{eq:primaloptimization} can be rewritten as
\begin{equation}\label{eq}
    \alpha := \underset{\rho}\min\,\underset{\sigma\in\mathcal{I}}\min\, {S}(\rho||\sum_{i}\Pi_i\sigma \Pi_i).
\end{equation}

Since our definition of coherence guarantees that the optimization problem satisfies the strong duality criterion, the primary optimization problem can be transformed into the dual problem \cite{zorzi2014minimum} by constructing a Lagrangian
\begin{equation}\label{}
    \mathcal{L} {(\rho, \lambda)}:={S}(\rho||\sum_{i}\Pi_i\rho \Pi_i)+\lambda[\mathrm{tr}(\rho M_1)-P(1|\rho)],
\end{equation}
where only one constraint in the primary problem exists, and the Lagrange multiplier $\vec{\lambda}$ is degraded to a scalar $\lambda$. The dual problem is given by
\begin{equation}\label{}
    \beta = \underset{\lambda}\max\,\underset{\rho}\min\, k\mathcal{L} {(\rho, \lambda)},
\end{equation}
where the factor $k=\ln(2)$. Strong duality implies that $\beta=\alpha$.

By introducing another function
\begin{equation}\label{}
    f(\rho,\sigma,\lambda):=S(\rho||\sum_{i}\Pi_i \sigma \Pi_i)+\lambda [\mathrm{tr}(\rho M_1)-P(1|\rho)],
\end{equation}
the dual problem can be expressed in the form of a three-level optimization problem
\begin{equation}\label{eq:hatbeta}
    \beta=\underset{\lambda}\max\,\underset{\rho}\min\,\underset{\sigma\in\mathcal{I}}\min\,kf(\rho,\sigma,\lambda).
\end{equation}

The two minimizations in Eq.(\ref{eq:hatbeta}) can be interchanged. We first solve $\underset{\rho\in\mathcal{C}}\min\,f(\rho,\sigma,\lambda)$, and acquire the unique solution and the optimal value \cite{zorzi2014minimum}
\begin{equation}\label{}
\begin{aligned}
    \rho^* &=\mathrm{exp}\left[-\mathbb{I}-\lambda M_1+\mathrm{ln}\left(\sum_{i}\Pi_i \sigma \Pi_i\right)\right] \\
    f(\rho^*,\sigma,\lambda) &=-\mathrm{tr}(\rho^*)-\lambda P(1|\rho^*).
\end{aligned}
\end{equation}
Using the Golden-Thompson inequality
\begin{equation}\label{}
    \mathrm{tr}[\mathrm{exp}(A+B)]\leq \mathrm{tr}[\mathrm{exp}(A) \mathrm{exp}(B)],
\end{equation}
we finally obtain
\begin{equation}\label{}
    \beta\geq \underset{\lambda}\max\,k[-||\sum_{i}\Pi_i R(\lambda)\Pi_i||-\lambda \mathrm{tr}(\rho M_1)],
\end{equation}
where
\begin{equation}\label{}
    ||\sum_{i}\Pi_iR(\lambda)\Pi_i||=\underset{\sigma\in\mathcal{I}}\max\,\mathrm{tr}[\sum_{i}\Pi_i R(\lambda)\Pi_i\sigma],
\end{equation}
\begin{equation}\label{}
    R(\lambda)=\mathrm{exp}(-\mathbb{I}-\lambda M_1).
\end{equation}

\section{Simulations}\label{smsec:para}

\subsection{Parameter estimation with decoy state method}

Given an ideal single photon source, one can perform a perfect tomography of the two-dimensional POVM for the measurement device, and directly apply
\begin{equation}\label{eq:targetfunc}
\max_\lambda[-||\sum_i \Pi_i \exp(-\mathbb{I}-\lambda M_1) \Pi_i ||-\lambda \mathrm{tr}(\rho M_1)]
\end{equation}
to calculate a lower bound of coherence. In practice, the source is a phase-randomized weak coherent state with an intensity of $\mu$, regarded as a Poisson distributed mixed state
\begin{equation}
\rho=e^{-\mu}\sum_n \frac{\mu^n}{n!}\ket{n}\bra{n},
\end{equation}
and thus the tomography cannot be accurate.

We use decoy state method to estimate the parameters of POVM, $a_0$, $n_x$, $n_y$ and $n_z$.
For an arbitrary state $j \in \{\ket{0}\bra{0},\ket{1}\bra{1},\ket{+}\bra{+},\ket{+i}\bra{+i},\rho\}$, prepared as a signal state $\mu$ or a decoy state $\nu$, we record the non-detection and double click events to be `0' and the probability of output `1' is
\begin{equation}\label{eq:pnu}
\begin{aligned}
p_\mu(1|j)&=e^{-\mu}\sum_{n=0} \frac{\mu^n}{n!} p_n(1|j) \\
p_\nu(1|j)&=e^{-\nu}\sum_{n=0} \frac{\nu^n}{n!} p_n(1|j),
\end{aligned}
\end{equation}
where $p_1(1|j)$ is the conditional probability of ideal single-photon source, corresponding to the ideal tomography result, and $p_0(1|j)=p_d$ is the dark count rate.

Similar to the parameter $Y_1$ in decoy state quantum key distribution (QKD), the lower and upper bounds of $p_1(1|j)$ can be estimated as \cite{ma2005practical}
\begin{equation}\label{eq:lowerbound}
\begin{aligned}
 \frac{\mu}{\mu\nu-\nu^2}&\left(p_\nu(1|j)e^\nu-p_\mu(1|j)e^\mu \frac{\nu^2}{\mu^2}-\frac{\mu^2-\nu^2}{\mu^2}p_d\right) \\& \le p_1(1|j) \\
  &\le \frac{p_\nu(1|j)}{\nu e^{-\nu}},
 \end{aligned}
\end{equation}
where the parameters of $p_\mu(1|j)$ and $p_\nu(1|j)$ are directly measured in the experiment.

We further consider the statistical fluctuations of $p_\mu(1|j)$ and $p_\nu(1|j)$. The number of runs for each type of state is $N \eta_j $, in which $\eta_j$ is the proportion of each type state and $\sum_j \eta_j=1$. The proportion of choosing signal state is $p_s$. The number of successful detections of a certain state is
\begin{equation}\label{eq:detectionnumber}
\begin{aligned}
M_{\mu,j}&=N\eta_jp_s p_\mu(1|j) \\
M_{\nu,j}&=N\eta_j(1-p_s)p_\nu(1|j).
\end{aligned}
\end{equation}
Therefore, the fluctuations of successful detections are
\begin{equation}\label{eq:detectionfluc}
\begin{aligned}
M_{\mu,j}^U&=M_{\mu,j}+n_\sigma\sqrt{M_{\mu,j}}	\\
M_{\mu,j}^L&=M_{\mu,j}-n_\sigma\sqrt{M_{\mu,j}} \\
M_{\nu,j}^U&=M_{\nu,j}+n_\sigma\sqrt{M_{\nu,j}} \\
M_{\nu,j}^L&=M_{\nu,j}-n_\sigma\sqrt{M_{\nu,j}},
\end{aligned}
\end{equation}
where we assume that the fluctuations follow a Gaussian distribution and $n_\sigma$ is set to be $3.89$ in our calculation, corresponding to a failure probability of $10^{-4}$.

Then the upper and lower bounds of $p_\mu(1|j)$ and $p_\nu(1|j)$ are calculated as
\begin{equation}\label{eq:sf}
\begin{aligned}
	p^U_{\mu}(1|j)&=\frac{M_{\mu,j}^U}{N\eta_j p_s} \\
	p^L_{\mu}(1|j)&=\frac{M_{\mu,j}^L}{N\eta_j p_s} \\
	p^U_{\nu}(1|j)&=\frac{M_{\nu,j}^U}{N\eta_j (1-p_s)} \\
	p^L_{\nu}(1|j)&=\frac{M_{\nu,j}^L}{N\eta_j (1-p_s)}.
\end{aligned}
\end{equation}
Combining Eq.~\eqref{eq:lowerbound} and Eq.~\eqref{eq:sf}, the lower and upper bounds after considering the statistical fluctuations are
\begin{equation}\label{eq:lowerandupperbound}
\begin{aligned}
p^L_{1}(1|j)&=\frac{\mu}{\mu\nu-\nu^2}\left(p^L_{\nu}(1|j)e^\nu-p^U_{\mu}(1|j)e^\mu\frac{\nu^2}{\mu^2}-\frac{\mu^2-\nu^2}{2\mu^2}p_d\right)	 \\
p^U_{1}(1|j)&=\frac{p^U_{\nu}(1|j)}{\nu e^{-\nu}}.
\end{aligned}
\end{equation}

Finally, we obtain the constraints for the optimization problem of Eq.~\eqref{eq:targetfunc} as
\begin{equation}\label{eq:decoybound}
p^L_{1}(1|j) \leq \mathrm{tr}(M_1 \rho_j) \leq p^U_{1}(1|j).
\end{equation}

\subsection{Intensity optimization and simulations}
Before the experiment, we need to optimize the intensities of signal and decoy states, i.e., $\mu$ and $\nu$. $p(1|j)$ with a $Y$ measurement (the actual measurement in the experiment) can be estimated as
\begin{equation}\label{eq:simu}
\begin{aligned}
p_{\mu}(1\big| \ket{0}\bra{0})&=1-(1-p_d)e^{-\eta \mu/2} \\
p_{\mu}(1\big| \ket{1}\bra{1})&=1-(1-p_d)e^{-\eta \mu/2} \\
p_{\mu}(1\big| \ket{+}\bra{+})&=1-(1-p_d)e^{-\eta \mu/2} \\
p_{\mu}(1\big| \ket{+i}\bra{+i})&=1-(1-p_d)e^{-\eta \mu} \\
p_{\nu}(1\big| \ket{0}\bra{0})&=1-(1-p_d)e^{-\eta \nu/2} \\
p_{\nu}(1\big| \ket{0}\bra{0})&=1-(1-p_d)e^{-\eta \nu/2} \\
p_{\nu}(1\big| \ket{+}\bra{+})&=1-(1-p_d)e^{-\eta \nu/2} \\
p_{\nu}(1\big| \ket{+i}\bra{+i})&=1-(1-p_d)e^{-\eta \nu}.
\end{aligned}
\end{equation}

By substituting Eq.~\eqref{eq:simu} into the Eq.~\eqref{eq:lowerbound} and combining Eq.~\eqref{eq:detectionfluc} and Eq.~\eqref{eq:sf}, we can calculate the corresponding lower and upper bounds. The target function is
\begin{equation}\label{eq:optmunu}
\begin{aligned}
f&=(p^U_{\mu}(1\big| \ket{0}\bra{0})-p^L_{\mu}(1\big| \ket{0}\bra{0}))^2 \\
&+(p^U_{\mu}(1\big| \ket{1}\bra{1})-p^L_{\mu}(1\big| \ket{1}\bra{1}))^2 \\
&+(p^U_{\mu}(1\big| \ket{+}\bra{+})-p^L_{\mu}(1\big| \ket{+}\bra{+}))^2 \\
&+(p^U_{\mu}(1\big| \ket{+i}\bra{+i})-p^L_{\mu}(1\big| \ket{+i}\bra{+i}))^2,
\end{aligned}
\end{equation}
with constraints of
\begin{equation}\label{eq:optmunuconstraint}
\begin{aligned}
p^L_{\mu}(1\big| \ket{0}\bra{0}) & \leq p^U_{\mu}(1\big| \ket{0}\bra{0})\\
p^L_{\mu}(1\big| \ket{1}\bra{1}) & \leq p^U_{\mu}(1\big| \ket{1}\bra{1})\\
p^L_{\mu}(1\big| \ket{+}\bra{+}) & \leq p^U_{\mu}(1\big| \ket{+}\bra{+})\\
p^L_{\mu}(1\big| \ket{+i}\bra{+i})& \leq  p^U_{\mu}(1\big| \ket{+i}\bra{+i})\\
 \nu &\leq \mu.
\end{aligned}
\end{equation}

Given the transmittance $\eta$, dark count rate $p_d$, number of total runs $N$, proportion of each type of state $\eta_j$, proportion of preparing signal states $p_s$ and the Gaussian fluctuation parameter $n_\sigma$, we can find the optimized $\mu_{opt}$ and $\nu_{opt}$ with Eq.~\eqref{eq:optmunu} and \eqref{eq:optmunuconstraint}. Then we can simulate the experimental results by substituting $\mu_{opt}$ and $\nu_{opt}$ into Eq.~\eqref{eq:simu} and further calculate $p^L_{1}(1|j)$ and $p^U_{1}(1|j)$, $j \in \{\ket{0}\bra{0},\ket{1}\bra{1},\ket{+}\bra{+},\ket{+i}\bra{+i},\rho\}$ according to Eq.~\eqref{eq:detectionnumber}-\eqref{eq:lowerandupperbound}. Finally, we find the coherence lower bound in Eq.~\eqref{eq:targetfunc} with constraints Eq.~\eqref{eq:decoybound}, and the lower bound of coherence with respect to $\eta$ can be simulated.

\subsection{Comparison with non-decoy case}
In order to show the advantage in precision of decoy state method, we perform a comparison with the non-decoy case where the source only prepares states with an intensity of $\mu$. The target function is still Eq.~\eqref{eq:targetfunc} while the constraint is looser than the decoy state case. We first optimize $\mu$ to make the constraints as tight as possible. The upper and lower bound of $p(1|j)$ without decoy state method is given by
\begin{equation}
\begin{aligned}
p^{\prime L}_{1}(1|j)&=\max\left\{0,\frac{p^L_{\mu}(1|j)-e^{-\mu}p_d-(1-e^{-\mu}-\mu e^{-\mu})}{\mu e^{-\mu}}\right\}	 \\
p^{\prime U}_{1}(1|j)&=\min\left\{1,\frac{p^U_{\mu}(1|j)}{\mu e^{-\mu}}\right\},
\end{aligned}
\end{equation}
where $p^L_{\mu}(1|j)$ and $p^U_{\mu}(1|j)$ are calculated in Eq.~\eqref{eq:sf}. We minimize
\begin{equation}
f^\prime=\sum_{j=1}^4 (p^{\prime U}_{1}(1|j)-p^{\prime L}_{1}(1|j))^2
\end{equation}
and find the optimized value $\mu_{opt}$. By optimizing Eq.~\eqref{eq:targetfunc} with constraints
\begin{equation}\label{eq:nondecoybound}
p^{\prime L}_{1}(1|j) \leq \mathrm{tr}(M_1 \rho_j) \leq p^{\prime U}_{1}(1|j),
\end{equation}
the lower bound of coherence with respect to $\eta$ can be simulated.

\section{Experiment}
\subsection{Intensity optimization and experimental coherence witness results}

The total transmission efficiency in the experiment is $\sim$ $4.86\%$, and the corresponding optimal intensity values are $\mu=0.529$ and $\nu=0.057$. With the experimental results of the conditional probabilities $p_\mu(1|j)$ and $p_\nu(1|j)$, $(j=1,2,3,4,5)$, we optimize Eq.~\eqref{eq:targetfunc} with constraints Eq.~\eqref{eq:decoybound}. The optimization result of coherence lower bound is $C^L=0.25$, corresponding to a random number generation rate of $R=kC^L \mu \eta=320$ kbps, where $k$ is the system clock rate.

\begin{figure*}[tbp]
\centering
\includegraphics[width=16.5 cm]{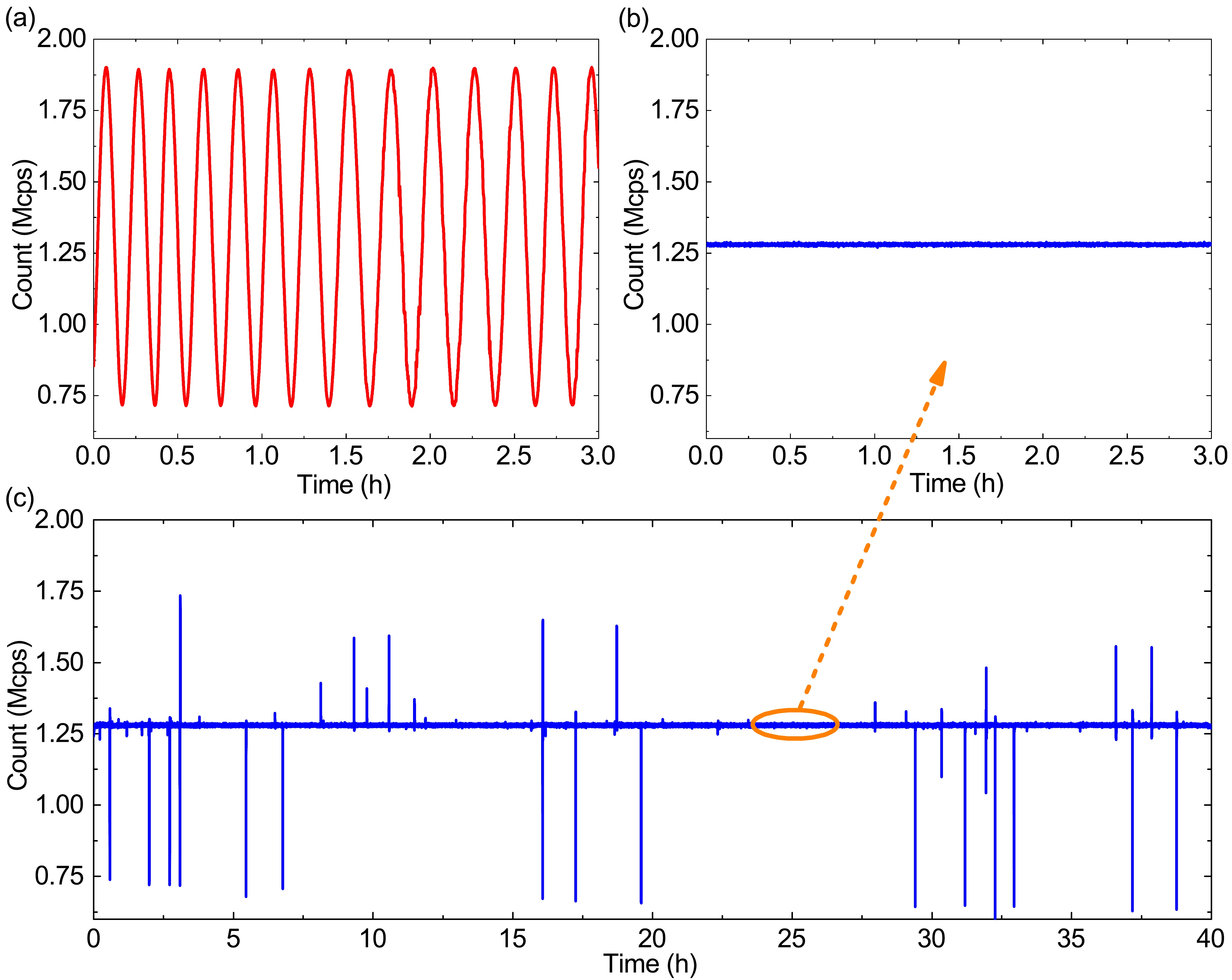}
\caption{A comparison of recorded count rate without (a) and with (b) phase stabilization over 3 hours. (c) Long-term phase stability over 40 hours.}
\label{figS1}%
\end{figure*}

\subsection{Phase stabilization of the interferometers}

In the experiment, maintaining the phase stabilization of two interferometers is the key technology for the time-bin encoding system. We apply an active feedback approach for the implementation of phase stability. The detection signals of SPAD are used to drive a counter (Keysight 53220A) via a 1:2 buffer. The counter is connected with a computer,
in which a proportional-integral-derivative(PID) algorithm is implemented. The PID program sends feedback signals to a high-voltage module (HVM, Thorlabs MDT694B) to precisely tune the phase shifter (PS) in real-time with a feedback frequency of 1 Hz. Then, the phase difference between the two interferometers is regulated.

Fig.~\ref{figS1}(a) and Fig.~\ref{figS1}(b) show a comparison of measured count rate without and with phase stabilization, respectively, from which one can clearly observe that the active feedback technology changes the count rate from a periodic oscillation to a straight line. We also test the long-term phase stability of the system over 40 hours, as
shown in Fig.~\ref{figS1}(c). The perpendicular lines are due to the reset process of active feedback. The HVM has a voltage range of 0-100 $V$. During the process of
active feedback, the initial output voltage of HVM is set at half of the range, and then the PID program increases or decreases the output voltage continuously to stabilize the count rate. After a long-time operation, the calculated output voltage may exceed the range. In such a case, the output voltage of HVM is reset to the initial value, and thus the sudden fluctuation of count rate is caused.

\subsection{Test results of prepared quantum states}
\begin{figure*}[tbp]
\centering
\includegraphics[width=16.5 cm]{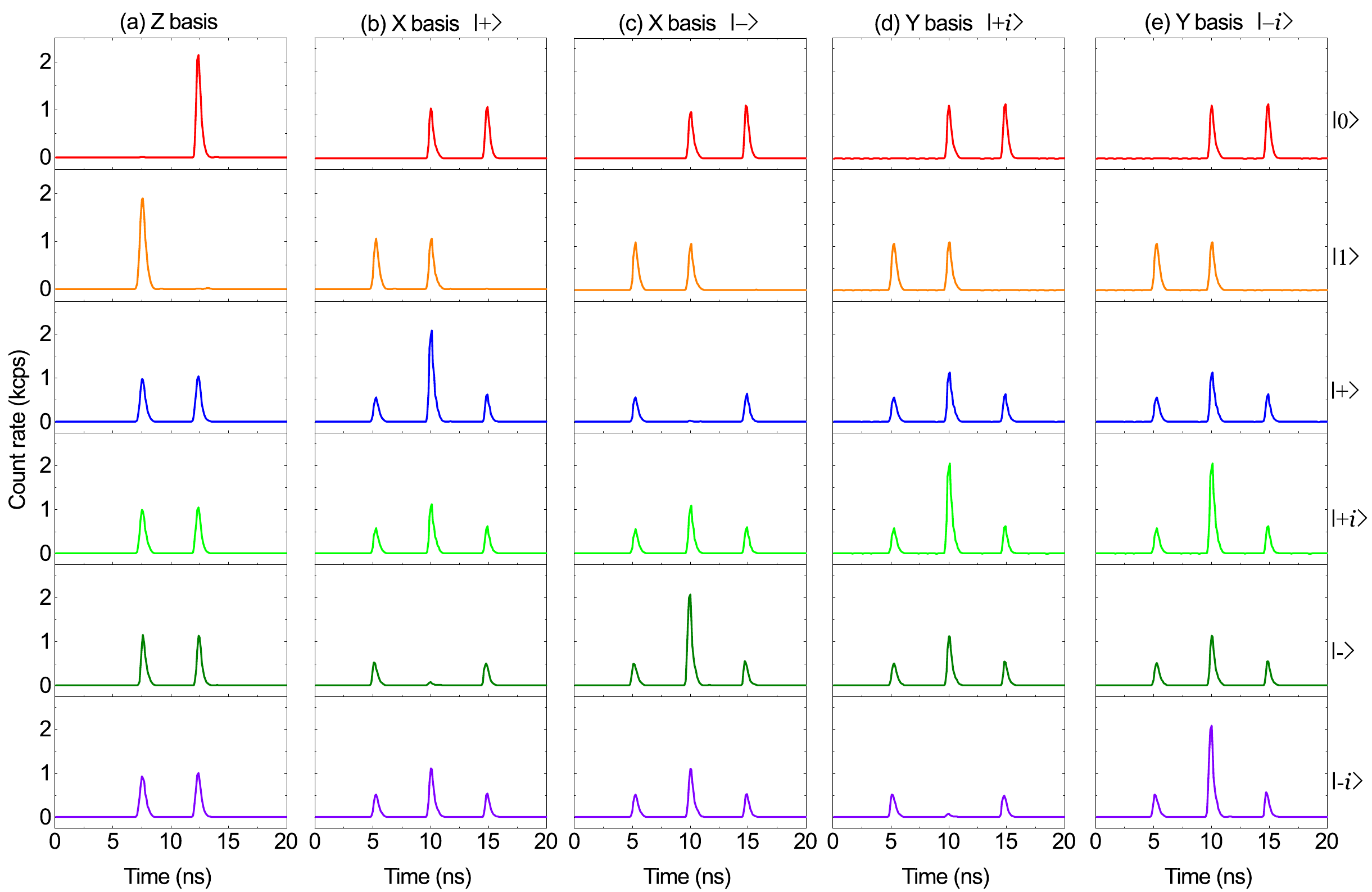}
\caption{Count rate distribution of the prepared quantum states measured in $X$, $Y$ and $Z$ bases using SPAD and TDC.}
\label{fig3}%
\end{figure*}

In the experiment, the quantum states of $\ket{0}$, $\ket{1}$, $\ket{+}$, $\ket{+i}$, $\ket{-}$ and $\ket{-i}$ are prepared and verified. Typical count rate distributions of the six time-bin states measured in $X$, $Y$, and $Z$ bases using SPAD and TDC are plotted in Fig.~\ref{fig3}.
To implement the $Z$ basis measurement, PM2 and the interferometer in the measurement part are not used. As shown in Fig.~\ref{fig3}(a), two time-bin pulses, i.e., an early one and a late one, are created by the unbalanced interferometer in the source part. When the early (late) pulse is removed by the AM, the state of $\ket{0}$ ($\ket{1}$) is prepared. When both of the pulses are attenuated to half by the AM and meanwhile the relative phase between two pulses is set as 0, $\frac{\pi}{2}$, $\pi$ or $\frac{3\pi}{2}$ by PM1, the corresponding state of $\ket{+}$, $\ket{+i}$, $\ket{-}$ or $\ket{-i}$ is prepared.
For $X$ ($Y$) basis measurement, the relative phase between two pulse is set as 0 ($\frac{\pi}{2}$) by PM2 and the count rate distributions of six time-bin states in basis $\ket{+}$, $\ket{-}$, $\ket{+i}$ and $\ket{-i}$ are shown in Fig.~\ref{fig3}(b), Fig.~\ref{fig3}(c), Fig.~\ref{fig3}(d) and Fig.~\ref{fig3}(e), respectively.

\begin{figure}[tbp]
\begin{minipage}[b]{1\linewidth}
\centering
\includegraphics[width=8.5 cm]{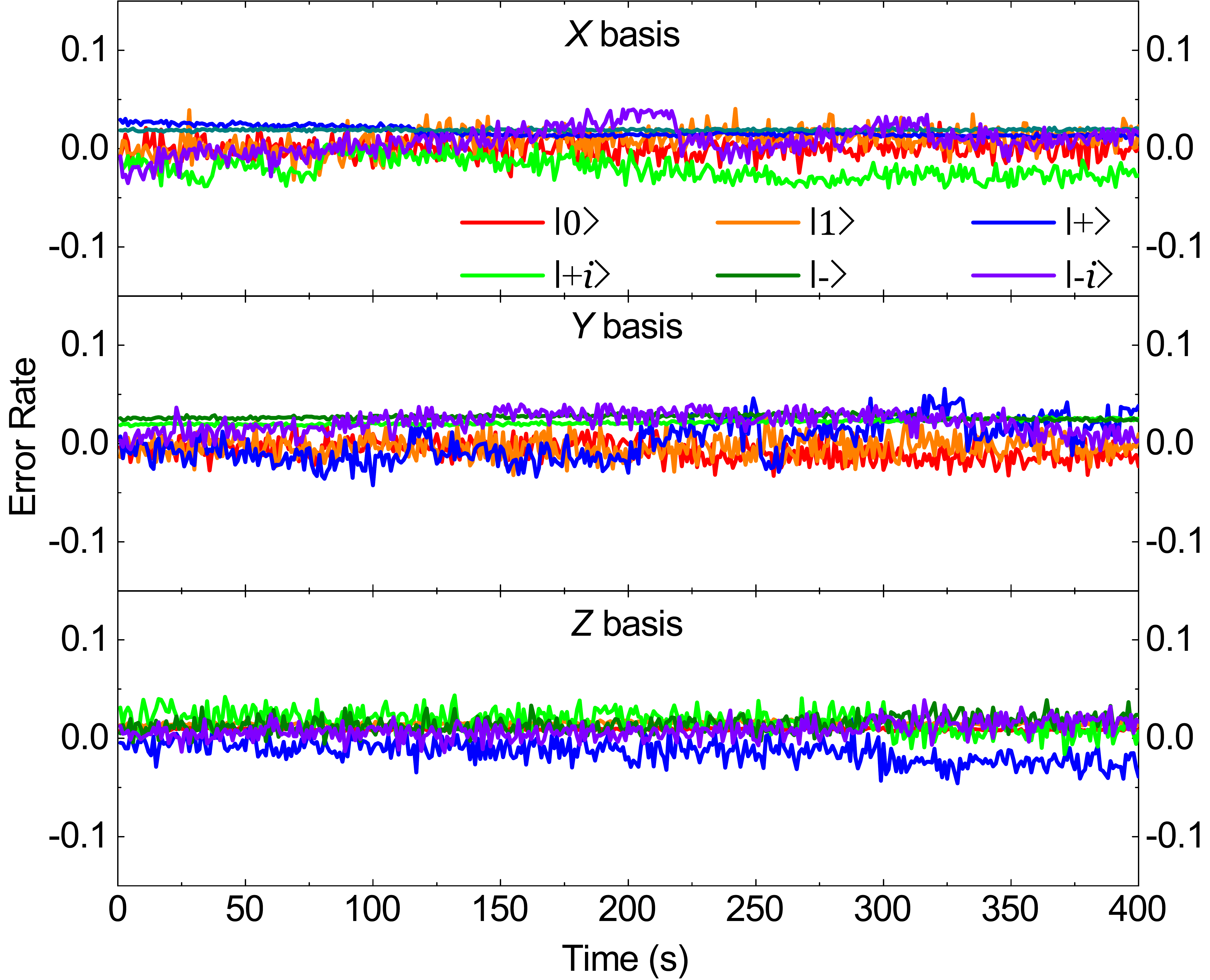}
\end{minipage}
~\\
\begin{minipage}[b]{1\linewidth}
\centering
\centering
\tabcolsep0.1in
\begin{tabular}{c|ccc}
  \hline
  \multirow{2}{*}{Quantum state}   & \multicolumn{3}{c}{Error rate}   \\
%  \cline{2-4}
                                &$X$ basis  & $Y$ basis  & $Z$ basis        \\
  \hline
        $\ket{0}$               & $0.03\%$  & $0.74\%$   & $0.90\%$         \\
        $\ket{1}$               & $0.83\%$  & $0.34\%$   & $1.55\%$         \\
        $\ket{+}$               & $1.72\%$  & $0.21\%$   & $1.37\%$         \\
        $\ket{+i}$              & $2.08\%$  & $1.23\%$   & $1.74\%$         \\
        $\ket{-}$               & $1.90\%$  & $2.70\%$   & $1.44\%$         \\
        $\ket{-i}$              & $0.81\%$  & $2.16\%$   & $0.85\%$         \\
  \hline
\end{tabular}
\end{minipage}
\caption{The fluctuations over 400 s (upper figure) and average values (lower table) of error rates after projecting the prepared quantum states in $X$, $Y$ and $Z$ bases.}\label{fig4}
\end{figure}

Further, we measure the error rates of the prepared states after the projection in $X$, $Y$, and $Z$ bases, respectively. The fluctuations of error rates over 400 s, and the average error rate values are shown in the upper figure and the lower table of Fig.~\ref{fig4}, respectively. Low values and slight fluctuations of error rates indicate the accuracy and stability of the quantum state preparation.

\subsection{Control experiment}
Here, we design and perform a control experiment using the same experimental setup. Different from the MDICW experiment, the unknown quantum state $\rho$ is replaced by a mixed state $\rho'$, which is an equal mixture of $\ket{+i}$ and $\ket{-i}$. In the control experiment, the four time-bin quantum states, $\ket{0}$, $\ket{1}$, $\ket{+}$, $\ket{+i}$, and the mixed state, $\rho'$, with intensities of $\mu$ and $\nu$, are randomly sent to the untrusted $Y$-basis measurement site. In total, $3.32 \times 10^{7}$ quantum states are sent to perform coherence witness.

\begin{table}[tbp]
\centering
\tabcolsep0.05in
\caption{Measurement tomography results in the case of unknown mixture information for the mixed state $\rho'$.
\label{tableS2}}
\begin{tabular}{lc|ccc}
  \hline
  \multicolumn{2}{c}{Test state}               & Amount   & Counts of `1' & Probability           \\
  \hline
  \multirow{4}{*}{Signal state}& $\ket{0}$     & 2047374  & 25488         & $1.24\times10^{-2}$   \\
                               & $\ket{1}$     & 2047692  & 24589         & $1.20\times10^{-2}$   \\
                               & $\ket{+}$     & 2044872  & 23269         & $1.14\times10^{-2}$   \\
                               & $\ket{+i}$    & 2048129  & 44997         & $2.20\times10^{-2}$   \\
                               & $\rho'$       & 8193886  & 94428         & $1.15\times10^{-2}$   \\
%  \cline{2-5}
  \hline
  \multirow{4}{*}{Decoy state} & $\ket{0}$     & 2047939  & 2445          & $1.19\times10^{-3}$   \\
                               & $\ket{1}$     & 2048117  & 2360          & $1.15\times10^{-3}$   \\
                               & $\ket{+}$     & 2047777  & 2243          & $1.10\times10^{-3}$   \\
                               & $\ket{+i}$    & 2048045  & 4325          & $2.11\times10^{-3}$   \\
                               & $\rho'$       & 8194169  & 9360          & $1.14\times10^{-3}$   \\
  \hline
\end{tabular}
\end{table}

\begin{table}[tbp]
\centering
\tabcolsep0.05in
\caption{Measurement tomography results in the case of unknown mixture information for the mixed state $\rho'$.
\label{tableS3}}
\begin{tabular}{lc|ccc}
  \hline
  \multicolumn{2}{c}{Test state}                    & Amount   & Counts of `1' & Probability           \\
  \hline
  \multirow{4}{*}{Signal state}& $\ket{0}$          & 2047374  & 25488         & $1.24\times10^{-2}$   \\
                               & $\ket{1}$          & 2047692  & 24589         & $1.20\times10^{-2}$   \\
                               & $\ket{+}$          & 2044872  & 23269         & $1.14\times10^{-2}$   \\
                               & $\ket{+i}$         & 2048129  & 44997         & $2.20\times10^{-2}$   \\
                               & $\rho', \ket{+i}$    & 4097561  & 91409         & $2.23\times10^{-2}$   \\
                               & $\rho', \ket{-i}$    & 4096325  & 3019          & $7.37\times10^{-4}$   \\
%  \cline{2-5}
  \hline
  \multirow{4}{*}{Decoy state} & $\ket{0}$          & 2047939  & 2445          & $1.19\times10^{-3}$   \\
                               & $\ket{1}$          & 2048117  & 2360          & $1.15\times10^{-3}$   \\
                               & $\ket{+}$          & 2047777  & 2243          & $1.10\times10^{-3}$   \\
                               & $\ket{+i}$         & 2048045  & 4325          & $2.11\times10^{-3}$   \\
                               & $\rho', \ket{+i}$    & 4095845  & 8475          & $1.14\times10^{-3}$   \\
                               & $\rho', \ket{-i}$    & 4098324  & 885           & $2.16\times10^{-4}$   \\
  \hline
\end{tabular}
\end{table}

In the case that the random mixture information for the state $\rho'$ is unknown, the measurement tomography results are listed in Table~\ref{tableS2}. As a calculated result, no coherence is witnessed.
When the random mixture information is known, i.e., the information that each sent pulse for the state $\rho'$ is $\ket{+i}$ or $\ket{-i}$ is provided,
the measurement tomography results are listed in Table~\ref{tableS3}. By applying the evaluation method of coherence witness, the coherence of only the state $\ket{+i}$ ($\ket{-i}$) is quantified with a lower bound of $0.0285$ ($0.1279$) per detected signal state. Therefore, the coherence is clearly witnessed in such a scenario. From the comparative results, one can conclude that our MDICW scheme can well witness the coherence of quantum states.

\subsection{Randomness generation}
As a direct application, the quantified coherence of an unknown state can be extracted as quantum random numbers.

The whole setup can be regarded as an MDICW-QRNG, in which the source part is switched between coherence witness mode and randomness generation mode while the measurement part is switched between $X$ and $Y$ bases.
In coherence witness mode, unknown state $\rho$ and four test states with intensities of $\mu$ or $\nu$ are randomly sent and $Y$ basis measurement is performed, in order to quantify the coherence.
In randomness generation mode, unknown signal state $\rho$ with intensities of $\mu$ is sent and $X$ basis measurement is performed such that raw random bits of `0' or `1' are generated.

In the experiment, a certain amount of random numbers are stored in the FPGA ahead as a random seed, which is used for state preparation. In order to gain output randomness more than input randomness, the number of experiment rounds used for the randomness generation mode is much larger than that for the coherence witness mode. In the randomness quantification, we need to consider statistical fluctuations due to the finite data size, which lies in two aspects of our protocol. One is the bias in the tomography result caused by statistical fluctuation, which has been well addressed in Eq.~\eqref{eq:detectionfluc} and \eqref{eq:sf}. The other is the randomness quantification by coherence in the finite data size case. That is, the relative entropy of coherence can only be used to quantify randomness in the asymptotic limit. We ignore the second fluctuation and leave it for future investigation. Here, we use the QRNG part as a simple demonstration of the application of MDICW. In summary, we directly use the lower bound of the relative entropy of coherence to quantify the output randomness.

\begin{table}[bp]
\tabcolsep0.02in
\caption{\label{TabS1} Typical NIST test results of the final random data with a size of 1 Gbits. The $p$-value and the proportion are set as 0.01 and 0.98, respectively.}
\begin{tabular}{l|lcl}
\hline
\hline
Statistical test &P-value &Proportion &Result\\
\hline
Frequency                  &0.4616   &0.992  &Pass\\
Block Frequency            &0.7439   &0.992  &Pass\\
Cumulative Sum             &0.3273   &0.993  &Pass\\
Runs                       &0.8343   &0.989  &Pass\\
Longest Run                &0.6931   &0.988  &Pass\\
Rank                       &0.4226   &0.989  &Pass\\
FFT                        &0.3221   &0.982  &Pass\\
Non Overlapping Template   &0.1189   &0.989  &Pass\\
Overlapping Template       &0.1672   &0.989  &Pass\\
Universal                  &0.5503   &0.993  &Pass\\
Approximate Entropy        &0.1230   &0.990  &Pass\\
Random Excursions          &0.1896   &0.989  &Pass\\
Random Excursions Variant  &0.0999   &0.990  &Pass\\
Serial                     &0.2812   &0.990  &Pass\\
Linear Complexity          &0.1001   &0.988  &Pass\\
\hline
\hline
\end{tabular}
\end{table}

In each round, $2^{32}$ quantum states in total including $10\times2^{15}$ quantum states for performing coherence witness are sent.
For each state in coherence witness mode, 3 random bits are used to determine the state for preparation and another 1 bit is used to determine the intensity of the state.
The detection information (`0' or `1') of each state is recorded.
Therefore, each round consumes 1152 Kb of random numbers while produces 4 Gb of raw data.
To bound the coherence accurately, the MDICW-QRNG process is performed for 100 rounds in total, so that $\sim$ 115 Mb of random numbers are consumed and 400 Gb of raw data are produced.
As a result, the coherence of the unknown state $\rho$ is quantified with a lower bound of $0.25$ per detected signal state.

In the MDICW-QRNG implementation, the parameters including the channel loss of 13.13 dB, the mean photon number of the unknown quantum state $\mu=0.529$, and the coherence lower bound of $0.25$ per detected signal state, correspond to a min-entropy of $6.4 \times 10^{-3}$ bits per pulse.
Since the system clock rate is $50$ MHz, the generation rate of MDICW-QRNG reaches $320$ kbps.
For randomness extraction, a Toeplitz-matrix hash function is applied~\cite{Ma13} and more than $2.5$ Gbits random numbers are finally obtained.
In order to verify the quality of the final random bits, the standard NIST statistical tests~\cite{NIST} are applied. Table~\ref{TabS1} shows that 1 Gbits final random numbers pass all the test items.

\end{document}